# Conductivity-Like Gilbert Damping due to Intraband Scattering in Epitaxial Iron

Behrouz Khodadadi[1], Anish Rai[2,3], Arjun Sapkota[2,3], Abhishek Srivastava[2,3], Bhuwan Nepal[2,3], Youngmin Lim[1], David A. Smith[1], Claudia Mewes[2,3], Sujan Budhathoki[2,3], Adam J. Hauser[2,3], Min Gao[4], Jie-Fang Li[4], Dwight D. Viehland[4], Zijian Jiang[1], Jean J. Heremans[1], Prasanna V. Balachandran[5,6], Tim Mewes[2,3], Satoru Emori[1]*

[1] *Department of Physics, Virginia Tech, VA 24061, U.S.A*

[2] *Department of Physics and Astronomy, University of Alabama, Tuscaloosa, AL 35487, USA*

[3] *Center for Materials for Information Technology (MINT), University of Alabama, Tuscaloosa, AL 35487, U.S.A.*

[4] *Department of Material Science and Engineering, Virginia Tech, Blacksburg, VA 24061, U.S.A.*

[5] *Department of Material Science and Engineering, University of Virginia, Charlottesville, VA 22904, U.S.A.*

[6] *Department of Mechanical and Aerospace Engineering, University of Virginia, Charlottesville, VA 22904, U.S.A.*

*email: semori@vt.edu

**Confirming the origin of Gilbert damping by experiment has remained a challenge for many decades, even for simple ferromagnetic metals. In this Letter, we experimentally identify Gilbert damping that increases with decreasing electronic scattering in epitaxial thin films of pure Fe. This observation of conductivity-like damping, which cannot be accounted for by classical eddy current loss, is in excellent quantitative agreement with theoretical predictions of Gilbert damping due to intraband scattering. Our results resolve**

**the longstanding question about a fundamental damping mechanism and offer hints for engineering low-loss magnetic metals for cryogenic spintronics and quantum devices.**

Damping determines how fast the magnetization relaxes towards the effective magnetic field and plays a central role in many aspects of magnetization dynamics [1,2]. The magnitude of viscous Gilbert damping governs the threshold current for spin-torque magnetic switching and auto-oscillations [3,4], mobility of magnetic domain walls [5,6], and decay lengths of diffusive spin waves and superfluid-like spin currents [7,8]. To enable spintronic technologies with low power dissipation, there is currently much interest in minimizing Gilbert damping in thin films of magnetic materials [9–13], especially ferromagnetic metals [14–23] that are compatible with conventional device fabrication schemes. Despite the fundamental and technological importance of Gilbert damping, its physical mechanisms in various magnetic materials have yet to be confirmed by experiment.

Gilbert damping is generally attributed to spin-orbit coupling that ultimately dissipates the energy of the magnetic system to the lattice [1,2]. Kambersky's torque correlation model [24] qualitatively captures the temperature dependence of damping in some experiments [25–28] by partitioning Gilbert damping into two mechanisms due to spin-orbit coupling, namely interband and intraband scattering mechanisms, each with a distinct dependence on the electronic momentum scattering time $\tau_e$. For the interband scattering mechanism where magnetization dynamics can excite electron-hole pairs across different bands, the resulting Gilbert damping is "resistivity-like" as its magnitude scales with $\tau_e^{-1}$, i.e., increased electronic scattering results in higher damping [29,30]. By contrast, the intraband scattering mechanism is typically understood through the breathing Fermi surface model [31], where electron-hole pairs are excited in the same band, yielding "conductivity-like" Gilbert damping that scales with $\tau_e$, i.e., reduced electronic scattering results in higher damping.

Conductivity-like Gilbert damping was reported experimentally more than 40 years ago in bulk crystals of pure Ni and Co at low temperatures, but surprisingly not in pure Fe [25]. The apparent absence of conductivity-like damping in Fe has been at odds with many theoretical predictions that intraband scattering should dominate at low temperatures [32–38], although some theoretical studies have suggested that intraband scattering may be absent altogether in pure metals [39,40]. To date, no experimental work has conclusively addressed the role of intraband scattering in pure Fe[1]. There thus remains a significant gap in the fundamental understanding of damping in one of the simplest ferromagnetic metals. Intrinsic conductivity-like Gilbert damping in Fe is also technologically relevant, since minimizing damping in ferromagnetic metals at low temperatures is crucial for cryogenic superconducting spintronic memories [41,42] and quantum information transduction schemes [43,44].

In this Letter, we experimentally demonstrate the presence of conductivity-like Gilbert damping due to intraband scattering in epitaxial thin films of body-centered-cubic (BCC) Fe. By combining broadband ferromagnetic resonance (FMR) measurements with characterization of structural and transport properties of these model-system thin films, we show that conductivity-like Gilbert damping dominates at low temperatures in epitaxial Fe. These experimental results agree remarkably well with the magnitude of Gilbert damping derived from first-principles calculations [32,33,36], thereby providing evidence for intraband scattering as a key mechanism for Gilbert damping in pure BCC Fe. Our experiment thus resolves the longstanding question regarding the origin of damping in the prototypical ferromagnetic metal. Our results also confirm that – somewhat counterintuitively – disorder can partially suppress intrinsic damping at low temperatures in ferromagnetic metals, such that optimally disordered films may be well suited for cryogenic spintronic and quantum applications [41–44].

---

[1] Ref. [36] includes experimental data that suggest the presence of conductivity-like Gilbert damping in an ultrathin Fe film, although no detailed information is given about the sample and the experimental results deviate considerably from the calculations. An earlier study by Rudd *et al.* also suggests an increase in Gilbert damping with decreasing temperature [27], but quantification of the Gilbert damping parameter in this experiment is difficult.

Epitaxial BCC Fe thin films were sputter deposited on (001)-oriented $MgAl_2O_4$ (MAO) and MgO single crystal substrates. The choices of substrates were inspired by the recent experiment by Lee *et al*. [20], where epitaxial growth is enabled with the [100] axis of a BCC Fe-rich alloy oriented 45º with respect to the [100] axis of MAO or MgO. MAO with a lattice parameter of $a_{MAO}/(2\sqrt{2}) = 0.2858$ nm exhibits a lattice mismatch of less than 0.4% with Fe ($a_{Fe} \approx 0.287$ nm), whereas the lattice mismatch between MgO ($a_{MgO}/\sqrt{2} = 0.2978$ nm) and Fe is of the order 4%. Here, we focus on 25-nm-thick Fe films that were grown simultaneously on MAO and MgO by confocal DC magnetron sputtering [45]. In the Supplemental Material [45], we report on additional films deposited by off-axis magnetron sputtering.

We verified the crystalline quality of the epitaxial Fe films by X-ray diffraction, as shown in Fig. 1(a-c). Only (00X)-type peaks of the substrate and film are found in each 2θ-ω scan, consistent with the single-phase epitaxial growth of the Fe films. The 2θ-ω scans reveal a larger amplitude of film peak for MAO/Fe, suggesting higher crystalline quality than that of MgO/Fe. Pronounced Laue oscillations, indicative of atomically smooth film interfaces, are observed around the film peak of MAO/Fe, whereas they are absent for MgO/Fe. The high crystalline quality of MAO/Fe is also evidenced by its narrow film-peak rocking curve with a FWHM of only 0.02°, comparable to the rocking curve FWHM of the substrate[2]. By contrast, the film-peak rocking curve of MgO/Fe has a FWHM of 1°, which indicates substantial mosaic spread in the film due to the large lattice mismatch with the MgO substrate.

Results of 2θ-ω scans for different film thicknesses [45] suggest that the 25-nm-thick Fe film may be coherently strained to the MAO substrate, consistent with the smooth interfaces and minimal mosaic spread of MAO/Fe. By contrast, it is likely that 25-nm-thick Fe on MgO is relaxed to accommodate the large film-substrate lattice mismatch. Static magnetometry provides further evidence that Fe is strained on MAO and relaxed on MgO [45]. Since strained MAO/Fe and relaxed MgO/Fe exhibit distinct crystalline quality, as evidenced by an approximately 50 times narrower rocking FWHM for MAO/Fe, we have two model systems that enable experimental investigation of the impact of structural disorder on Gilbert damping.

The residual electrical resistivity also reflects the structural quality of metals. As shown in Fig. 1(d), the residual resistivity is 20% lower for MAO/Fe compared to MgO/Fe, which corroborates the lower defect density in MAO/Fe. The resistivity increases by nearly an order of magnitude with increasing temperature, reaching $1.1 \times 10^{-7}$ Ω m for both samples at room temperature, consistent with behavior expected for pure metal thin films.

We now examine how the difference in crystalline quality correlates with magnetic damping in MAO/Fe and MgO/Fe. Broadband FMR measurements were performed at room temperature up to 65 GHz with a custom spectrometer that employs a coplanar waveguide (center conductor width 0.4 mm) and an electromagnet (maximum field < 2 T). For each measurement at a fixed excitation frequency, an external bias magnetic field was swept parallel to the film plane along the [110] axis of Fe, unless otherwise noted. In the Supplemental

---

[2] The angular resolution of the diffractometer is 0.0068°.

Material [45], we show similar results with the field applied along the [110] and [100] axes of Fe; Gilbert damping is essentially isotropic within the film plane for our epitaxial Fe films, in contrast to a recent report of anisotropic damping in ultrathin epitaxial Fe [22].

Figure 2 shows that the peak-to-peak FMR linewidth $\Delta H_{pp}$ scales linearly with frequency $f$, enabling a precise determination of the measured Gilbert damping parameter $\alpha_{meas}$ from the standard equation,

$$\mu_0 \Delta H_{pp} = \mu_0 \Delta H_0 + \frac{2}{\sqrt{3}} \frac{\alpha_{meas}}{\gamma'} f, \qquad (1)$$

where $\Delta H_{pp,0}$ is the zero-frequency linewidth and $\gamma' = \gamma/2\pi \approx 29.5$ GHz/T is the reduced gyromagnetic ratio. Despite the difference in crystalline quality, we find essentially the same measured Gilbert damping parameter of $\alpha_{meas} \approx 2.3 \times 10^{-3}$ for MAO/Fe and MgO/Fe. We note that this value of $\alpha_{meas}$ is comparable to the lowest damping parameters reported for epitaxial Fe at room temperature [15–17]. Our results indicate that Gilbert damping at room temperature is insensitive to the strain state or structural disorder in epitaxial Fe.[3]

The measured damping parameter $\alpha_{meas}$ from in-plane FMR can generally include a contribution from non-Gilbert relaxation, namely two-magnon scattering driven by defects [46–49]. However, two-magnon scattering is suppressed when the film is magnetized *out-of-plane* [19,48]. To isolate any two-magnon scattering contribution to damping, we performed out-of-plane FMR measurements under a sufficiently large magnetic field (>4 T) for complete saturation of the Fe film, using a custom W-band shorted waveguide combined with a superconducting magnet. As shown in Fig. 2, the out-of-plane and in-plane FMR data yield the same slope and hence $\alpha_{meas}$ (Eq. 1) to within < 8%. This finding indicates that two-magnon scattering is negligible and that frequency-dependent magnetic relaxation is dominated by Gilbert damping in epitaxial Fe examined here.

---

[3] However, the crystallographic texture of Fe has significant impact on damping; for example, non-epitaxial Fe films deposited directly on amorphous $SiO_2$ substrates exhibit an order of magnitude wider linewidths, due to much more pronounced non-Gilbert damping (e.g., two-magnon scattering), compared to (001)-oriented epitaxial Fe films.

The insensitivity of Gilbert damping to disorder found in Fig. 2 can be explained by the dominance of the interband (resistivity-like) mechanism at room temperature, with phonon scattering dominating over defect scattering. Indeed, since MAO/Fe and MgO/Fe have the same room-temperature resistivity (Fig. 1(d)), any contributions to Gilbert damping from electronic scattering should be identical for both samples at room temperature. Moreover, according to our density functional theory calculations [45], the density of states of BCC Fe at the Fermi energy, $D(E_F)$, does not depend significantly on the strain state of the crystal. Therefore, in light of the recent reports that Gilbert damping is proportional to $D(E_F)$ [18,50,51], the different strain states of MAO/Fe and MgO/Fe are not expected to cause a significant difference in Gilbert damping.

However, since MAO/Fe and MgO/Fe exhibit distinct resistivities (electronic scattering times $\tau_e$) at low temperatures, one might expect to observe distinct temperature dependence in Gilbert damping for these two samples. To this end, we performed variable-temperature FMR measurements using a coplanar-waveguide-based spectrometer (maximum frequency 40 GHz, field < 2 T) equipped with a closed-cycle cryostat[4]. Figure 3(a,b) shows that $\alpha_{meas}$ is enhanced for both samples at lower temperatures. Notably, this damping enhancement with decreasing temperature is significantly greater for MAO/Fe. Thus, at low temperatures, we find a conductivity-like damping increase that is evidently more pronounced in epitaxial Fe with less structural disorder.

---

[4] The W-band spectrometer for out-of-plane FMR (Fig. 2) could not be cooled below room temperature due to its large thermal mass, limiting us to in-plane FMR measurements at low temperatures.

While this increased damping at low temperatures is reminiscent of intrinsic Gilbert damping from intraband scattering [31–38], we first consider other possible contributions. One possibility is two-magnon scattering [46–49], which we have ruled out at room temperature (Fig. 2) but could be present in our low-temperature in-plane FMR measurements. From Fig. 3(a,b), the zero-frequency linewidth $\Delta H_0$ (Eq. 1) – typically attributed to magnetic inhomogeneity – is shown to increase along with $\alpha_{\text{meas}}$ at low temperatures [45], which might point to the emergence of two-magnon scattering [48,49]. However, our mean-field model calculations (see Supplemental Material [45]) shows that $\Delta H_0$ correlates with $\alpha_{\text{meas}}$ due to interactions among different regions of the inhomogeneous film [52]. The increase of $\Delta H_0$ at low temperatures is therefore readily accounted for by increased Gilbert damping, rather than two-magnon scattering.

We are also not aware of any mechanism that enhances two-magnon scattering with decreasing temperature, particularly given that the saturation magnetization (i.e., dipolar interactions) is constant across the measured temperature range [45]. Moreover, the isotropic in-plane damping found in our study is inconsistent with typically anisotropic two-magnon scattering tied to the crystal symmetry of epitaxial films [46,47], and the film thickness in our study (e.g., 25 nm) rules out two-magnon scattering of interfacial origin [49]. As such, we conclude that two-magnon scattering does not play any essential role in our experimental observations.

Another possible contribution is dissipation due to classical eddy currents, which increases proportionally with the increasing conductivity $\sigma$ at lower temperatures. We estimate the eddy current contribution to the measured Gilbert damping with [15,53]

$$\alpha_{eddy} = \frac{\sigma}{12}\gamma\mu_0^2 M_s t_F^2, \qquad (2)$$

where $\mu_0 M_s \approx 2.0$ T is the saturation magnetization and $t_F$ is the film thickness. We find that eddy current damping accounts for only ≈20% (≈30%) of the total measured damping of MAO/Fe (MgO/Fe) even at the lowest measured temperature (Fig. 3(c)). Furthermore, as shown in the Supplemental Material [45], thinner MAO/Fe films, e.g., $t_F = 11$ nm, with negligible $\alpha_{eddy}$ still exhibit a significant increase in damping with decreasing temperature. Our results thus indicate a substantial contribution to conductivity-like Gilbert damping that is not accounted for by classical eddy current damping.

For further discussion, we subtract the eddy-current damping from the measured damping to denote the Gilbert damping parameter attributed to intrinsic spin-orbit coupling as $\alpha_{so} = \alpha_{meas} - \alpha_{eddy}$. To correlate electronic transport and magnetic damping across the entire measured temperature range, we perform a phenomenological fit of the temperature dependence of Gilbert damping with [26]

$$\alpha_{so} = c\frac{\sigma(T)}{\sigma(300\,K)} + d\frac{\rho(T)}{\rho(300\,K)}, \qquad (3)$$

where the conductivity-like (intraband) and resistivity-like (interband) terms are scaled by adjustable parameters $c$ and $d$, respectively. As shown in Fig. 4(a),(b), this simple phenomenological model using the experimental transport results (Fig. 1(d)) agrees remarkably well with the temperature dependence of Gilbert damping for both MAO/Fe and MgO/Fe.

Our findings that Gilbert damping can be phenomenologically partitioned into two distinct contributions (Eq. 3) are in line with Kambersky's torque correlation model. We compare our experimental results to first-principles calculations by Gilmore *et al.* [32,33] that relate electronic momentum scattering rate $\tau_e^{-1}$ and Gilbert damping through Kambersky's torque correlation model. We use the experimentally measured resistivity ρ (Fig. 1(d)) to convert the temperature to $\tau_e^{-1}$ by assuming the constant conversion factor $\rho\tau_e = 1.30\times10^{-21}$ $\Omega$ m s [33]. To account for the difference in electronic scattering time for the minority spin $\tau\downarrow$ and majority spin $\tau\uparrow$, we take the calculated curve from Gilmore *et al.* with $\tau\downarrow/\tau\uparrow = 4$ [33], which is close to the ratio of $D(E_F)$ of the spin-split bands for BCC Fe, e.g., derived from our density functional theory calculations [45]. For explicit comparison with Refs. [32,33], the Gilbert damping parameter in Fig. 4(c) is converted to the magnetic relaxation rate $\lambda = \gamma\alpha_{so}\mu_0 M_s$. The calculated prediction is in excellent quantitative agreement with our experimental results for both strained MAO/Fe and relaxed MgO/Fe (Fig. 4(c)), providing additional experimental evidence that intraband scattering predominately contributes to Gilbert damping at low temperatures.

We also compare our experimental results to a more recent first-principles calculation study by Mankovsky *et al.*, which utilizes the linear response formalism [36]. This approach does not rely on a phenomenological electronic scattering rate and instead allows for explicitly incorporating thermal effects and structural disorder. Figure 4(d) shows the calculated temperature dependence of the Gilbert damping parameter for BCC Fe with a small density of defects, i.e., 0.1% vacancies, adapted from Ref. [36]. We again find good quantitative agreement between the calculations and our experimental results for MAO/Fe. On the other hand, the Gilbert damping parameters at low temperatures for relaxed MgO/Fe are significantly below the calculated values. This is consistent with the reduction of intraband scattering due to enhanced electronic scattering (enhanced $\tau_e^{-1}$) from defects in relaxed MgO/Fe.

Indeed, significant defect-mediated electronic scattering may explain the absence of conductivity-like Gilbert damping for crystalline Fe in prior experiments. For example, Ref. [25] reports an upper limit of only a two-fold increase of the estimated Gilbert damping parameter from $T = 300$ K to 4 K. This relatively small damping enhancement is similar to that for MgO/Fe in our study (Fig. 4(b)), suggesting that intraband scattering may have been suppressed in Fe in Ref. [25] due to a similar degree of structural disorder to MgO/Fe. We therefore conclude that conductivity-like Gilbert damping from intraband scattering is highly sensitive to disorder in ferromagnetic metals.

More generally, the presence of defects in all real metals – evidenced by finite residual resistivity – ensures that the Gilbert damping parameter is finite even in the zero-temperature limit. This circumvents the theoretical deficiency of Kambersky's torque correlation model where Gilbert damping would diverge in a perfectly clean ferromagnetic metal at $T \rightarrow 0$ [39,40]. We also remark that a fully quantum mechanical many-body theory of magnetization dynamics yields finite Gilbert damping even in the clean, $T = 0$ limit [54].

In summary, we have demonstrated the dominance of conductivity-like Gilbert damping due to intraband scattering at low temperatures in high-quality epitaxial Fe. Our experimental results also validate the longstanding theoretical prediction of intraband scattering as an essential mechanism for Gilbert damping in pure ferromagnetic metals [32–38], thereby advancing the fundamental understanding of magnetic relaxation in real materials. Moreover, we have confirmed that, at low temperatures, a magnetic metal with imperfect crystallinity can exhibit lower Gilbert damping (spin decoherence) than its cleaner counterpart. This somewhat counterintuitive finding suggests that magnetic thin films with optimal structural or chemical disorder may be useful for cryogenic spintronic memories [41,42] and spin-wave-driven quantum information systems [43,44].

Acknowledgements

This research was funded in part by 4-VA, a collaborative partnership for advancing the Commonwealth of Virginia, as well as by the ICTAS Junior Faculty Award. A. Sapkota and C. Mewes would like to acknowledge support by NSF-CAREER Award No. 1452670, and A. Srivastava would like to acknowledge support by NASA Award No. CAN80NSSC18M0023. We thank M. D. Stiles, B. K. Nikolic, and F. Mahfouzi for helpful discussions on theoretical models for computing Gilbert damping, as well as R. D. McMichael for his input on the mean-field modeling of interactions in inhomogeneous ferromagnetic films.

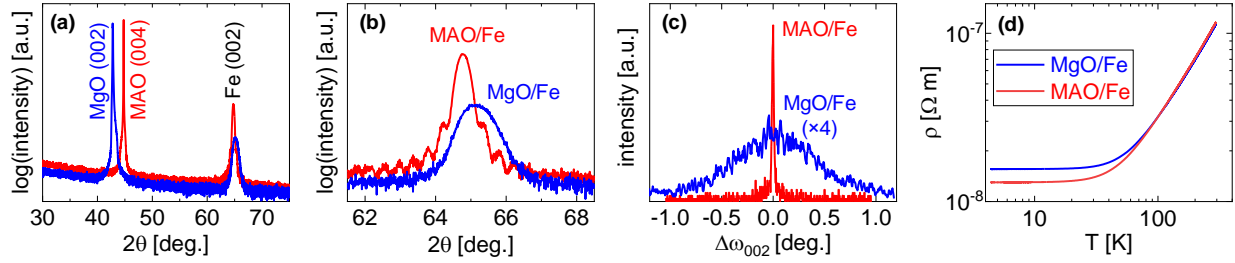

Figure 1. (a,b) 2θ-ω X-ray diffraction scans of MAO/Fe and MgO/Fe (a) over a wide angle range and (b) near the BCC Fe (002) film peak. (c) Rocking curve scans about the film peak. (d) Temperature dependence of resistivity plotted on a log-log scale.

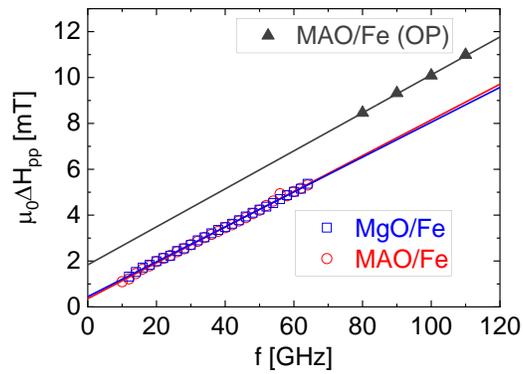

Figure 2. Frequency dependence of FMR linewidth $\Delta H_{pp}$ for MAO/Fe and MgO/Fe at room temperature. Linewidths measured under in-plane field are shown as open symbols, whereas those measured under out-of-plane (OP) field are shown as filled symbols.

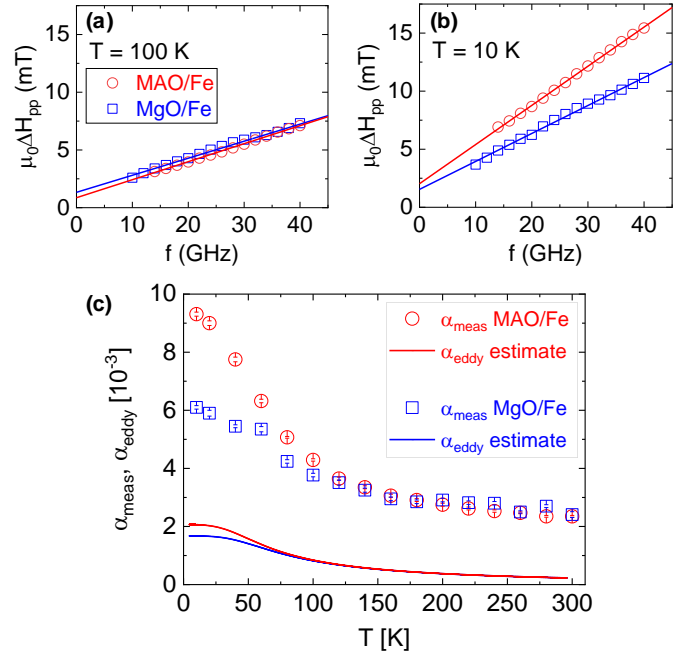

Figure 3. (a,b) Frequency dependence of FMR linewidth for MAO/Fe and MgO/Fe at (a) $T = 100$ K and (b) $T = 10$ K. (c) Temperature dependence of measured Gilbert damping parameter $\alpha_{meas}$ and estimated eddy-current damping parameter $\alpha_{eddy}$.

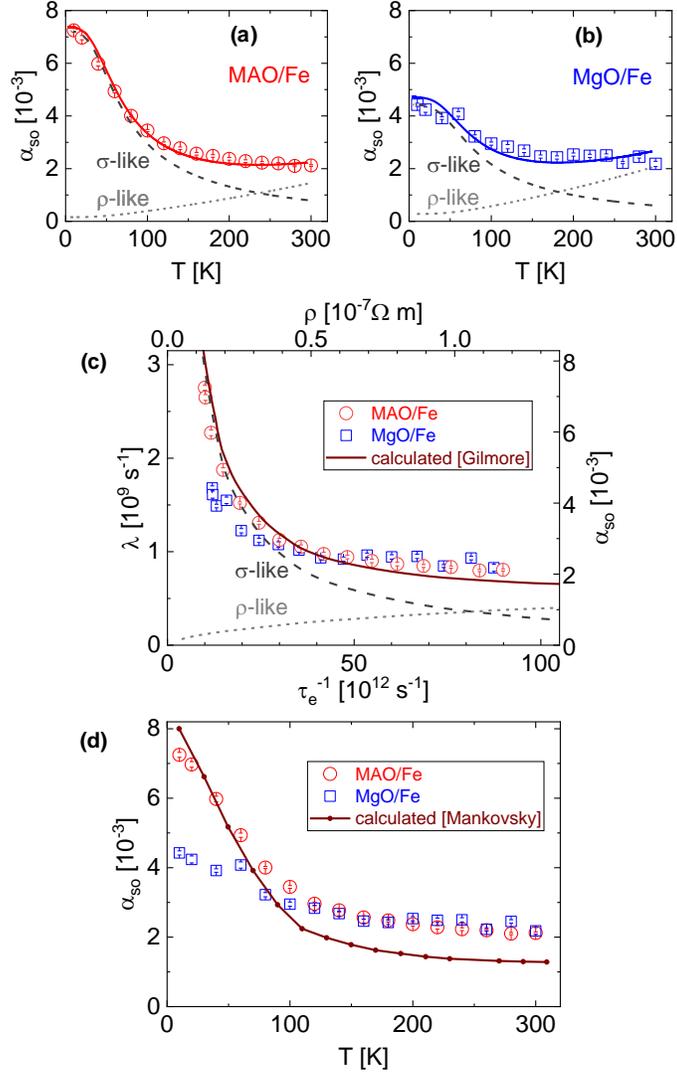

Figure 4. (a,b) Temperature dependence of the spin-orbit-induced Gilbert damping parameter $\alpha_{so}$, fit phenomenologically with the experimentally measured resistivity for (a) MAO/Fe and (b) MgO/Fe. The dashed and dotted curves indicate the conductivity-like and resistivity-like contributions, respectively; the solid curve represents the fit curve for the total spin-orbit-induced Gilbert damping parameter. (c,d) Comparison of our experimental results with calculated Gilbert damping parameters by (c) Gilmore *et al*. [32,33] and (d) Mankovsky *et al*. [36].